# Differentiable eigendecomposition-free RCWA for full-tensor anisotropic photonics

ENBO YANG[1, †], QIANG SONG[2, 3, †], WEIWEI CAI[1, *]

[1]Key Lab of Education Ministry for Power Machinery and Engineering, School of Mechanical Engineering, Shanghai Jiao Tong University, Shanghai 200240, China.

[2]Greater Bay Area Institute for Innovation, Hunan University, Guangzhou 511300, China

[3]Goertek Omnilights NanoOptics Co., Ltd., Shanghai 201413, China

[†]These authors contributed to the work equally.

* cweiwei@sjtu.edu.cn

**Abstract**

Full-tensor anisotropy transforms rigorous coupled-wave analysis (RCWA) into a large, fully coupled non-Hermitian eigenproblem, making eigendecomposition expensive and difficult to differentiate. We introduce a differentiable, eigendecomposition-free RCWA framework for spatially patterned media with fully coupled permittivity tensors, using boundary fields rather than internal eigenmodes as the layer representation. A boundary-value cascade constructs scattering operators directly from these fields, enabling automatic differentiation and efficient GPU execution. Benchmarks against finite-element and transfer-matrix solutions show close agreement in scattering responses, while automatic-differentiation gradients agree with finite differences and enable topology optimization. At 529 Fourier harmonics, layer construction is 22.8 times faster than conventional eigendecomposition on the same GPU. Our framework offers a general computational route toward scalable forward modeling and inverse design in anisotropic photonic systems.

## 1. INTRODUCTION

Periodic photonic structures incorporating anisotropic media offer degrees of freedom unavailable in isotropic systems, enabling polarization control, tunable wavefront manipulation, advanced displays, and flat optical devices [1–5]. Their optical response is governed not only by the unit-cell geometry but also by the spatial distribution and orientation of the permittivity tensors. Rotations of the principal axes introduce off-diagonal tensor components that couple nominally orthogonal polarization channels [6,7]. Extending electromagnetic modeling to spatially patterned, arbitrarily oriented tensors therefore changes the algebraic structure of the full-vector problem rather than simply increasing the number of material parameters. This distinction is particularly important in inverse design [8–11], where the electromagnetic problem must be solved repeatedly as both geometry and material tensors evolve.

Rigorous coupled-wave analysis (RCWA), also known as the Fourier modal method, is widely used for layered periodic structures [12–18]. Conventional RCWA resolves intralayer propagation through modal eigendecomposition. This approach is efficient for isotropic media, whose symmetries and partial field decoupling reduce the associated eigenproblem. Full-tensor anisotropy breaks these simplifications, producing a larger non-Hermitian eigensystem with stronger modal coupling and a more complex spectrum [6,7,15,16]. In the formulation considered here, the system matrix doubles in linear dimension, raising the nominal cubic cost of eigendecomposition by approximately eightfold at the same Fourier truncation. The resulting cost, together with the limited parallel scalability of dense eigendecomposition, makes full-tensor RCWA increasingly expensive on GPU architectures [19–22].

For photonic inverse design, differentiable forward solvers are increasingly important [23–25]. Several RCWA frameworks already support automatic differentiation for isotropic media using Lorentzian broadening to regularize singular eigengradients at modal degeneracies and thereby maintain finite numerical gradients [10,26] (Table 1). In fully anisotropic systems, however, stronger field coupling and more involved propagation-mode classification introduce additional challenges to gradient stability. Eigendecomposition-free formulations have therefore been explored in isotropic settings, replacing modal decomposition with operations better suited to parallel execution and automatic differentiation [19–22]. The central challenge in extending these methods to anisotropic media is not the inclusion of anisotropic material parameters, but the construction of a stable scattering formulation after the symmetries and structural simplifications of isotropic media are lost.

In this work, we develop an eigendecomposition-free RCWA framework for spatially patterned media with fully coupled permittivity tensors, supporting automatic differentiation and GPU acceleration (Table 1). We formulate anisotropic RCWA as a boundary-field problem and construct the layer scattering matrix with the boundary-value cascade method (BCM), avoiding both eigendecomposition and mode classification. At 529 Fourier harmonics, BCM achieves up to a 22.8-fold speedup over the conventional eigendecomposition-based approach. We validate the framework across diverse anisotropic structures, verify its automatic-differentiation gradients, and demonstrate free-form topology optimization. This boundary-value framework provides a scalable computational basis for differentiable modeling and inverse design of anisotropic photonic systems.

Table 1. Comparison of representative RCWA implementations on patterned periodic layers.

| **Framework solver path** | **Isotropic** | **Anisotropy** | **Automatic differentiation** | **Eigendecomposition-free acceleration** | **GPU Accele-ration** |
|---|---|---|---|---|---|
| S4 [13] | Yes | In-plane | No | No | No |
| RETICOLO [14] | Yes | Diagonal | No | No | No |
| rcwa_tf [10] | Yes | No | Yes | No | Yes |
| TORCWA [26] | Yes | No | Yes | No | Yes |
| FMMAX [27] | Yes | In-plane | Yes | No | Yes |
| GRCWA [28] | Yes | Diagonal | Yes | No | No |
| Meent [29] | Yes | Diagonal | Yes | No | Yes |
| TorchRDIT [20] | Yes | No | Yes | Yes | Yes |
| This work | Yes | Full-tensor | Yes | Yes | Yes |

## 2. METHOD

### *A. Overview of anisotropic RCWA*

We consider structures that are periodic in the transverse plane and piecewise layered along z. Without loss of generality, we assume that each layer contains spatially varying $3\times3$ relative permittivity and permeability tensors. The time dependence is exp(-jωt), and the magnetic field is normalized by the free-space impedance. Starting from Maxwell's equations, we expand the fields in the Fourier basis and apply Li's factorization rules [15,16]. Eliminating the longitudinal field components then yields a first-order system for the tangential field vector $\boldsymbol{\psi} = \begin{pmatrix} S_x & S_y & U_x & U_y \end{pmatrix}^T$ in the *i*-th layer

$$\frac{d\boldsymbol{\psi}}{d\tilde{z}} = \mathbf{M}^{(i)}\boldsymbol{\psi}, \tag{1}$$

where $\mathbf{M}^{(i)} = \begin{pmatrix} \boldsymbol{p} & \boldsymbol{P} \\ \boldsymbol{Q} & \boldsymbol{q} \end{pmatrix}$ is a 4 × 4 block matrix, with the complete expressions provided in Supplementary Material A. Conventional RCWA recasts Eq. (1) as the eigenvalue problem

$$\mathbf{M}^{(i)}\mathbf{V} = \boldsymbol{\lambda}\mathbf{V}. \tag{2}$$

In Eq. (2), the columns of $\mathbf{V}$ are the electromagnetic eigenmodes admitted by $\mathbf{M}^{(i)}$, and $\boldsymbol{\lambda}$ contains their complex propagation constants. The modes can be separated into forward- and backward-propagating components according to their propagation direction along *z*. In anisotropic media, coupling between field components prevents a simple symmetry-based separation of forward and backward modes [15,16]. These modes are classified by the direction of the longitudinal Poynting flux, whereas evanescent modes are classified by the sign of the imaginary propagation constant. After sorting, the tangential field inside the layer is expressed as:

$$\boldsymbol{\psi}(\tilde{z}) = \begin{pmatrix} \boldsymbol{V}_S^+ & \boldsymbol{V}_S^- \\ \boldsymbol{V}_U^+ & \boldsymbol{V}_U^- \end{pmatrix} \begin{pmatrix} e^{\boldsymbol{\lambda}^+\tilde{z}} & 0 \\ 0 & e^{\boldsymbol{\lambda}^-\tilde{z}} \end{pmatrix} \begin{pmatrix} \boldsymbol{c}^+ \\ \boldsymbol{c}^- \end{pmatrix} \tag{3}$$

where $\boldsymbol{c}$ represents the modal amplitudes. The boundary conditions on the two sides of the layer determine the scattering matrix of the *i*-th layer. The global S-matrix is then obtained by cascading the layer scattering matrices with the Redheffer star product, a standard procedure that is not repeated here [17,18].

This modal construction is useful when the internal field profile is required. For forward simulation and inverse design, however, the required quantity is usually the relation between incoming and outgoing fields at the layer boundaries. Computing and sorting the internal eigenmodes is then an avoidable intermediate step. The overhead is especially pronounced for full-tensor anisotropy, where all tangential field components are coupled and the layer operator is generally non-Hermitian. We therefore construct the boundary-value relation directly in the next section.

### *B. Scattering-matrix solution using boundary-value cascade method*

Our derivation builds on the bisection-doubling and incremental-storage principles of the precise integration method, as detailed in Supplementary Material B.1 [30–33]. The BCM provides the boundary-value relation without diagonalizing the layer operator or explicitly forming a modal basis. It builds the finite-thickness response by initializing a short interval and recursively combining adjacent intervals. We first rewrite Eq. (1) as

$$\begin{aligned}\frac{d\boldsymbol{S}}{d\tilde{z}} &= \boldsymbol{pS} + \boldsymbol{PU},\\ \frac{d\boldsymbol{U}}{d\tilde{z}} &= \boldsymbol{QS} + \boldsymbol{qU}.\end{aligned} \tag{4}$$

For the $i$-th layer, the normalized thickness is $\tilde{d}_i = jk_0 d_i$. Consider an interval $[\tilde{z}_a, \tilde{z}_b]$ within a layer. The tangential electromagnetic fields at its two sides can be written in an S-matrix-like port form

$$\begin{pmatrix}\boldsymbol{S}_b\\ \boldsymbol{U}_a\end{pmatrix} = \begin{pmatrix}\boldsymbol{A} & -\boldsymbol{B}\\ \boldsymbol{C} & \boldsymbol{D}\end{pmatrix}\begin{pmatrix}\boldsymbol{S}_a\\ \boldsymbol{U}_b\end{pmatrix}. \tag{5}$$

The subscripts $\boldsymbol{a}$ and $\boldsymbol{b}$ denote the left and right ports, respectively. The matrix $\boldsymbol{\Phi} = \begin{pmatrix}\boldsymbol{A} & -\boldsymbol{B}\\ \boldsymbol{C} & \boldsymbol{D}\end{pmatrix}$ is the coefficient matrix to be determined. Once $\boldsymbol{\Phi}$ is obtained, the corresponding scattering matrix can be constructed. As illustrated in Fig. 1, BCM partitions a thick layer into thin sublayers, solves the above boundary-value problem over a thin sublayer, and reconstructs the full-layer response through recursive bisection doubling.

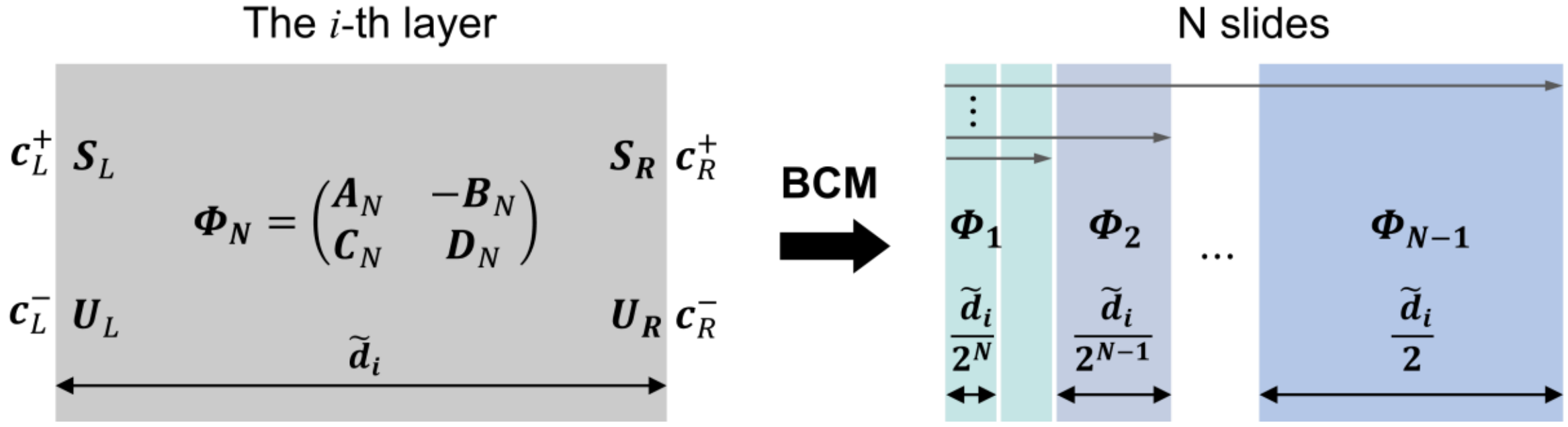

Fig. 1. Direct construction of the layer boundary relation. A thin-interval relation is initialized from a local Taylor expansion. Repeated interval combination doubles the represented thickness until the full layer is reached. The method cascades the coefficient matrix $\boldsymbol{\Phi}_n$ rather than a set of layer eigenmodes.

Specifically, the normalized layer thickness is divided into $2^N$ subintervals of equal length

$$\tilde{\tau} = \frac{\tilde{d}_i}{2^N}, \tag{6}$$

where $N$ is the BCM order. For a sufficiently small subinterval, the coefficient matrices are initialized by a fourth-order Taylor expansion

$$\begin{aligned}\boldsymbol{A}_1 &= \boldsymbol{I} + \boldsymbol{\alpha}_1\tilde{\tau} + \boldsymbol{\alpha}_2\tilde{\tau}^2 + \boldsymbol{\alpha}_3\tilde{\tau}^3 + \boldsymbol{\alpha}_4\tilde{\tau}^4 \equiv \boldsymbol{I} + \Delta\boldsymbol{A}_1\\ \boldsymbol{B}_1 &= \boldsymbol{\beta}_1\tilde{\tau} + \boldsymbol{\beta}_2\tilde{\tau}^2 + \boldsymbol{\beta}_3\tilde{\tau}^3 + \boldsymbol{\beta}_4\tilde{\tau}^4\\ \boldsymbol{C}_1 &= \boldsymbol{\gamma}_1\tilde{\tau} + \boldsymbol{\gamma}_2\tilde{\tau}^2 + \boldsymbol{\gamma}_3\tilde{\tau}^3 + \boldsymbol{\gamma}_4\tilde{\tau}^4\\ \boldsymbol{D}_1 &= \boldsymbol{I} + \boldsymbol{\delta}_1\tilde{\tau} + \boldsymbol{\delta}_2\tilde{\tau}^2 + \boldsymbol{\delta}_3\tilde{\tau}^3 + \boldsymbol{\delta}_4\tilde{\tau}^4 \equiv \boldsymbol{I} + \Delta\boldsymbol{D}_1.\end{aligned} \tag{7}$$

Combining Eqs. (4) and (5) yields the constraint equations satisfied by the above parameters, as detailed in Supplementary Material B.2. These constraints can then be solved to obtain the expressions for the unknown coefficients, which are provided in Supplementary Material C. Two adjacent equal intervals with matrix $\boldsymbol{\Phi}_n = \begin{pmatrix}\boldsymbol{A}_n & -\boldsymbol{B}_n\\ \boldsymbol{C}_n & \boldsymbol{D}_n\end{pmatrix}$ are then joined by eliminating their shared boundary fields. This gives the doubling recurrence

$$\begin{aligned} \boldsymbol{A}_{n+1} &= \boldsymbol{A}_n(\boldsymbol{I}+\boldsymbol{B}_n\boldsymbol{C}_n)^{-1}\boldsymbol{A}_n \\ \boldsymbol{B}_{n+1} &= \boldsymbol{B}_n+\boldsymbol{A}_n(\boldsymbol{I}+\boldsymbol{B}_n\boldsymbol{C}_n)^{-1}\boldsymbol{B}_n\boldsymbol{D}_n \\ \boldsymbol{C}_{n+1} &= \boldsymbol{C}_n+\boldsymbol{D}_n(\boldsymbol{I}+\boldsymbol{C}_n\boldsymbol{B}_n)^{-1}\boldsymbol{C}_n\boldsymbol{A}_n \\ \boldsymbol{D}_{n+1} &= \boldsymbol{D}_n(\boldsymbol{I}+\boldsymbol{C}_n\boldsymbol{B}_n)^{-1}\boldsymbol{D}_n. \end{aligned} \tag{8}$$

Repeated application of Eq. (8) doubles the represented thickness until the full layer is spanned. Since $\boldsymbol{A}$ and $\boldsymbol{D}$ differ only slightly from the identity on the initial subinterval, we store the increments to avoid loss of significance. Defining $\boldsymbol{G}_{An}=(I+\boldsymbol{B}_n\boldsymbol{C}_n)^{-1}$ and $\boldsymbol{G}_{Dn}=(I+\boldsymbol{C}_n\boldsymbol{B}_n)^{-1}$, the corresponding updates are

$$\begin{aligned} \Delta\boldsymbol{A}_{n+1} &= \left(\Delta\boldsymbol{A}_n-\frac{1}{2}\boldsymbol{B}_n\boldsymbol{C}_n\right)\boldsymbol{G}_{An}+\boldsymbol{G}_{An}\left(\Delta\boldsymbol{A}_n-\frac{1}{2}\boldsymbol{B}_n\boldsymbol{C}_n\right)+\Delta\boldsymbol{A}_n\boldsymbol{G}_{An}\Delta\boldsymbol{A}_n \\ \Delta\boldsymbol{D}_{n+1} &= \left(\Delta\boldsymbol{D}_n-\frac{1}{2}\boldsymbol{C}_n\boldsymbol{B}_n\right)\boldsymbol{G}_{Dn}+\boldsymbol{G}_{Dn}\left(\Delta\boldsymbol{D}_n-\frac{1}{2}\boldsymbol{C}_n\boldsymbol{B}_n\right)+\Delta\boldsymbol{D}_n\boldsymbol{G}_{Dn}\Delta\boldsymbol{D}_n \\ \boldsymbol{B}_{n+1} &= \boldsymbol{B}_n+(\boldsymbol{I}+\Delta\boldsymbol{A}_n)\boldsymbol{G}_{An}\boldsymbol{B}_n(\boldsymbol{I}+\Delta\boldsymbol{D}_n) \\ \boldsymbol{C}_{n+1} &= \boldsymbol{C}_n+(\boldsymbol{I}+\Delta\boldsymbol{D}_n)\boldsymbol{G}_{Dn}\boldsymbol{C}_n(\boldsymbol{I}+\Delta\boldsymbol{A}_n). \end{aligned} \tag{9}$$

After N doublings, the boundary-value relation across the full layer is

$$\begin{pmatrix}\boldsymbol{S}_R\\ \boldsymbol{U}_L\end{pmatrix}=\begin{pmatrix}\boldsymbol{I}+\Delta\boldsymbol{A}_N & -\boldsymbol{B}_N\\ \boldsymbol{C}_N & \boldsymbol{I}+\Delta\boldsymbol{D}_N\end{pmatrix}\begin{pmatrix}\boldsymbol{S}_L\\ \boldsymbol{U}_R\end{pmatrix}. \tag{10}$$

Eqs. (6)–(10) therefore advance the boundary-relation coefficient matrix from a local series initialization to the complete layer without constructing internal eigenmodes. Introducing a zero-thickness homogeneous gap then transforms this relation into the standard S-matrix for the ***i***-th layer

$$\boldsymbol{S}^{(i)}=\begin{bmatrix}-\boldsymbol{A}_N\boldsymbol{W}_0 & \boldsymbol{W}_0+\boldsymbol{B}_N\boldsymbol{V}_0\\ \boldsymbol{V}_0+\boldsymbol{C}_N\boldsymbol{W}_0 & \boldsymbol{D}_N\boldsymbol{V}_0\end{bmatrix}^{-1}\begin{bmatrix}\boldsymbol{A}_N\boldsymbol{W}_0 & -(\boldsymbol{W}_0-\boldsymbol{B}_N\boldsymbol{V}_0)\\ \boldsymbol{V}_0-\boldsymbol{C}_N\boldsymbol{W}_0 & \boldsymbol{D}_N\boldsymbol{V}_0\end{bmatrix}. \tag{11}$$

Here, $\boldsymbol{W}_0, \boldsymbol{V}_0$ are the electric- and magnetic-field eigenmodes in free space, respectively. The complete derivation from the boundary-value formulation to the scattering matrix is provided in Supplementary Material B.3. The scattering-matrix construction above avoids eigendecomposition of the large non-Hermitian matrices arising from the coupling of field components and polarization channels in anisotropic layers. Instead, the problem is recast as recursive updates of boundary relations using standard matrix operations. This formulation can reduce the computational cost and is naturally compatible with automatic differentiation, making it well suited for gradient-based inverse design and topology optimization of anisotropic photonic structures.

## 3. RESULTS

We evaluate the BCM-based RCWA formulation through four sets of numerical tests designed to examine distinct aspects of the method. We first quantify its accuracy, Fourier convergence, thickness dependence, and computational scaling using a tilted-axis lithium-niobate metasurface, where the off-diagonal tensor components couple the transverse and longitudinal electromagnetic fields. We then use an anisotropic multilayer with a semi-analytical Berreman transfer-matrix reference to independently assess polarization coupling and boundary matching. Volume liquid-crystal Bragg polarization gratings provide a more demanding test in which the material tensor varies continuously in both the transverse and longitudinal directions. Finally, we verify the automatic-differentiation gradients and demonstrate topology optimization of an anisotropic photonic structure.

Unless otherwise stated, the calculations set BCM order N=15, 529 truncated Fourier harmonics, and complex64 arithmetic with 32-bit real and imaginary components. The BCM solver is implemented in PyTorch and runs on an Intel Xeon Gold 6226R CPU and an NVIDIA GeForce RTX 4090 GPU. For brevity, the proposed method is denoted BCM and the conventional eigendecomposition-based formulation is denoted EIG.

### *A. Accuracy, convergence, and computational scaling*

We consider the tilted-axis lithium-niobate nanofin in Fig. 2(a) to test whether eliminating intralayer eigendecomposition compromises the accuracy of RCWA for a fully coupled anisotropic tensor. Here, the optical axis of lithium niobate is tilted in the *xz* plane. A nonzero tilt $\xi$ activates the *xz* and *zx* components of the permittivity tensor

$$\boldsymbol{\varepsilon}_{LN}(\xi) = \begin{bmatrix} \varepsilon_e\cos^2\xi + \varepsilon_o\sin^2\xi & 0 & (\varepsilon_e - \varepsilon_o)\sin\xi\cos\xi \\ 0 & \varepsilon_o & 0 \\ (\varepsilon_e - \varepsilon_o)\sin\xi\cos\xi & 0 & \varepsilon_e\sin^2\xi + \varepsilon_o\cos^2\xi \end{bmatrix}, \tag{12}$$

where $\varepsilon_o$ and $\varepsilon_e$ are the ordinary and extraordinary relative permittivities. We set $\xi = 45°$ and neglect dispersion for this test.

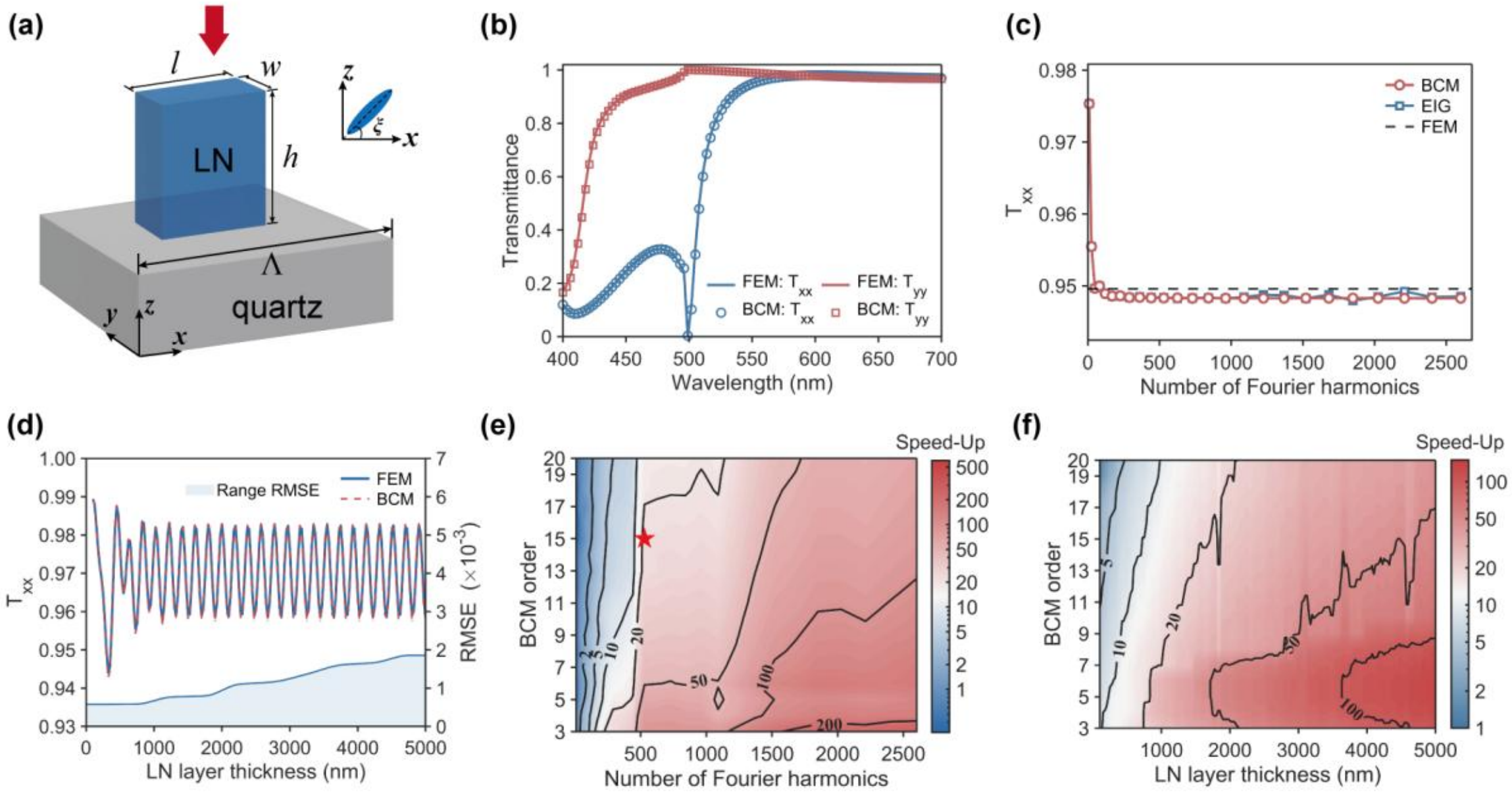

Fig. 2. Accuracy, convergence, and computational scaling for a tilted-axis lithium-niobate nanofin. (a) Geometry and crystal-axis orientation. The period is 340 nm, the height is 300 nm, the length is 200 nm, and the width is 100 nm. The nanofin is supported by fused silica and illuminated from air at normal incidence. We use $\varepsilon_o$ = 5.46 and $\varepsilon_e$ = 5.06 without material dispersion. (b) Co-polarized transmission spectra from 400 to 700 nm. (c) Convergence of BCM and EIG with Fourier harmonic count at 550 nm. FEM is the independent reference. (d) Transmission through lithium-niobate layers from 100 nm to 5 μm. Shading gives the RMSE within consecutive 500 nm intervals. (e) BCM speedup over EIG for constructing the single-layer S matrix on the same GPU. (f) BCM speedup over FEM for end-to-end solution on the same CPU. For BCM, a Fourier truncation of 361 harmonics is applied.

Fig. 2(b) compares the co-polarized transmittances obtained from BCM and FEM over 400–700 nm. The two methods reproduce the same spectral response, including the feature near 500 nm, with RMSE values below 0.5% over the scanned range. The agreement indicates that BCM retains the full-tensor electromagnetic coupling without mode decomposition inside the patterned layer. We next examine Fourier convergence at 550 nm. As the number of truncated Fourier harmonics increases, BCM and EIG converge toward nearly the same solution, while BCM exhibits higher numerical stability at high harmonic numbers [Fig. 2(c)]. The converged BCM and FEM values are 0.9496 and 0.9483, respectively, corresponding to a difference of approximately 0.1%

Another concern for eigendecomposition-free methods is the accumulation of error over optically thick layers [22]. Fig. 2(d) illustrates a scan of the lithium-niobate thickness from 100 nm to 5 µm. BCM reproduces the oscillatory FEM response throughout this range. The interval-wise RMSE increases gradually with optical thickness but remains below 0.2% throughout the tested range. These results show that our framework remains accurate over the tested thickness range without requiring the complete physical layer to satisfy a thin-layer approximation.

The computational advantage of BCM becomes more pronounced as the Fourier basis expands. Fig. 2(e) compares the construction of the lithium-niobate-layer scattering matrix using BCM and EIG on the same GPU. Although BCM incurs greater overhead at low harmonic counts, these truncations are insufficient for converged simulations. As the number of Fourier harmonics increases, the relative cost of EIG rises more rapidly, and under a representative truncation with 529 harmonics, BCM achieves a 22.8-fold speedup (red star). We also compare the performance of BCM and FEM over a range of layer thicknesses and BCM orders on the same CPU (Intel Core i9-14900K), as shown in Fig. 2(f). The observed speedup grows with layer thickness, consistent with the different longitudinal discretizations used by the BCM and FEM. While increasing the BCM order adds computational overhead, substantial speedups are maintained over a broad range of thicknesses.

### *B. Anisotropic-multilayer benchmark*

To isolate the treatment of anisotropic field coupling and boundary matching from transverse discretization, we next consider the multilayer structure in Fig. 3(a). The principal axis of lithium niobate is rotated within the *xy* plane, producing polarization coupling under oblique incidence. Because the structure is laterally homogeneous, only the zeroth Fourier harmonic is retained, allowing the BCM solution to be compared directly with the independent 4 × 4 Berreman transfer-matrix method (TMM) [34].

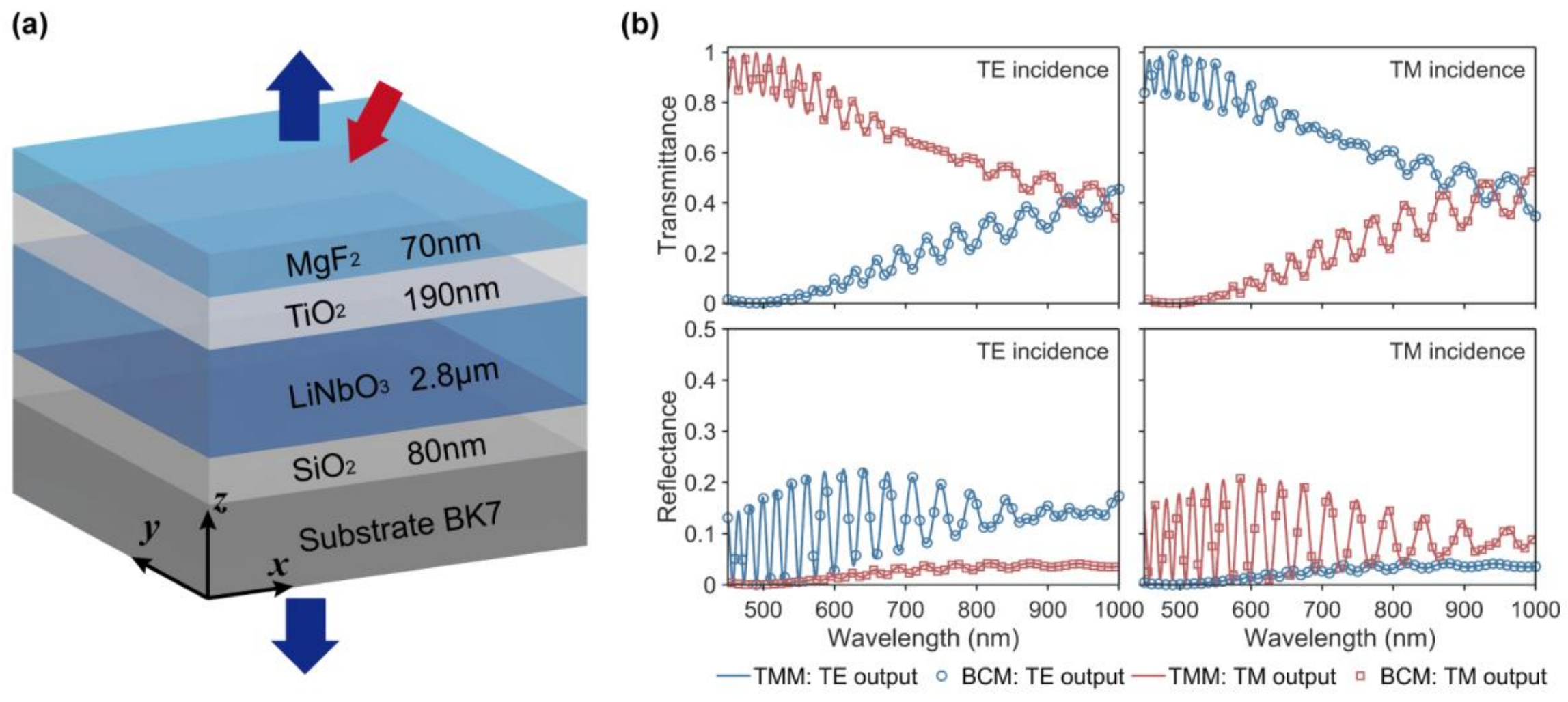

Fig. 3. Polarization-resolved benchmark for an anisotropic multilayer. (a) The stack contains air, 70 nm $MgF_2$, 190 nm $TiO_2$, 2.8 μm $LiNbO_3$, 80 nm $SiO_2$, and BK7. The refractive indices of $MgF_2$, $TiO_2$, $SiO_2$, and BK7 are 1.38, 2.40, 1.46, and 1.52, respectively. The lithium-niobate optical axis has an in-plane azimuth of 45 degrees. TE- and TM-polarized waves are incident from air at 30 degrees. (b) Transmittance and reflectance from 450 to 1000 nm. Solid lines denote the TMM, and open symbols denote BCM. A BCM order of 20 is used because the anisotropic layer is optically thick.

This configuration tests propagation through an optically thick anisotropic layer. Over the spectrum, the optical thickness of the lithium-niobate layer reaches several wavelengths, while the rotated anisotropy produces substantial conversion between TE and TM polarization channels. As shown in Fig. 3(b), BCM and TMM agree across all co-polarized and cross-polarized transmission and reflection channels, with RMSE values below 1%. Across the full spectral range, the maximum deviation of the total reflectance plus total transmittance from unity is $3.72 \times 10^{-3}$, consistent with energy conservation. This benchmark removes two ambiguities associated with discretized numerical solvers: zeroth-order truncation eliminates Fourier-convergence error, while the Berreman formulation provides an independent semi-analytical reference for the same anisotropic propagation problem. The agreement therefore shows that BCM preserves polarization coupling and boundary matching.

### *C. Volume liquid-crystal Bragg polarization gratings*

We next test whether the formulation remains applicable when the anisotropic tensor varies continuously in both the transverse and longitudinal directions. A liquid-crystal Bragg polarization grating provides such a case: the optical axis rotates periodically in the transverse plane and twists along the layer thickness [Fig. 4(a)], producing an inclined Bragg structure that selectively diffracts opposite circular-polarization states [35,36]. The orientation angle of the liquid crystal director is

$$\Phi(x,z) = \frac{\pi x}{\Lambda_{\mathrm{x}}} + \Omega z, \tag{13}$$

where $\Lambda_{\mathrm{x}}$ is the surface period and $\Omega$ is the longitudinal twist rate. We also take into account the out-of-plane tilt angle $\Theta$ of the optical axis. The director and relative-permittivity tensor are

$$\boldsymbol{u} = [\cos\Theta\cos\Phi, \cos\Theta\sin\Phi, \sin\Theta]^T,$$
$$\boldsymbol{\varepsilon} = \varepsilon_o \boldsymbol{I} + (\varepsilon_e - \varepsilon_o)\boldsymbol{u}\boldsymbol{u}^T. \quad (14)$$

Because the material tensor varies continuously along z, the structure is represented by a sequence of longitudinal slices. Within each slice, BCM calculates the scattering matrix from the local tensor distribution, and the slice scattering matrices are subsequently combined through the Redheffer star product [Fig. 4(b)]. Both transmissive and reflective gratings are considered.

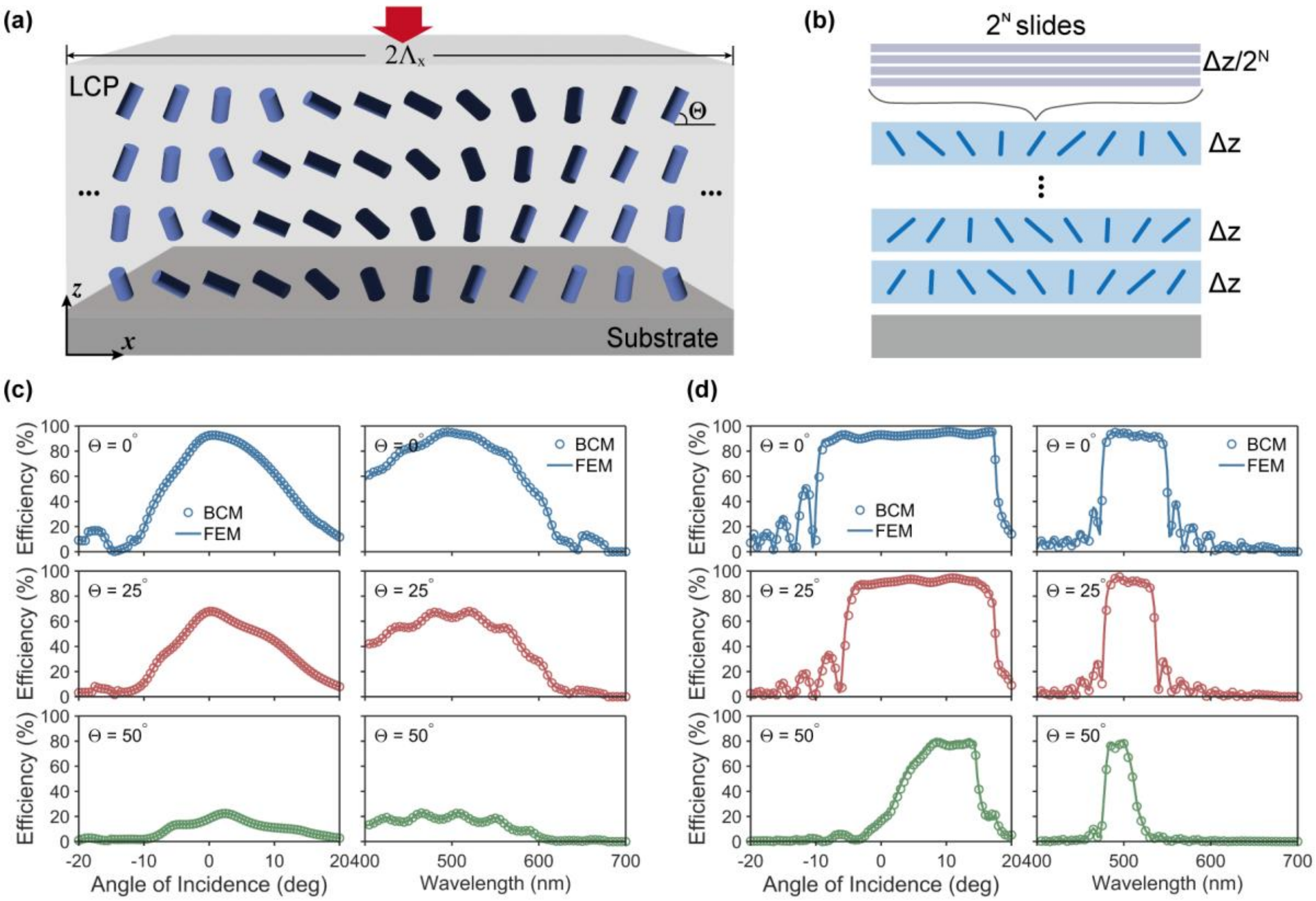

Fig. 4. Volume liquid-crystal Bragg polarization gratings. (a) Inclined director field and the full transverse period of the material tensor. The ordinary and extraordinary refractive indices are 1.525 and 1.775, the substrate index is 1.7, and $\Lambda_x$ = 400 nm. (b) Longitudinal slicing of the continuously varying tensor and BCM bisection within each slice. (c) Angular and spectral first-order diffraction from the 1 μm transmissive grating at Θ = 0, 25, and 50 degrees. The longitudinal twist rate is 217° μm⁻¹, and the grating is divided into 20 slices. (d) Corresponding responses of the 3 μm reflective grating. The longitudinal twist rate Ω = 929° μm⁻¹, and the grating is divided into 100 slices. Solid lines denote FEM, and open symbols denote BCM.

Figs. 4(c) and 4(d) compare the BCM and FEM responses for out-of-plane tilt angles $\Theta = 0°, 25°, \text{and } 50°$. For the transmissive grating, the angular and spectral responses agree well at all three tilt angles, with RMSE values below 1%. Both methods reproduce the reduction in diffraction efficiency as the tilt angle increases. For the reflective grating, BCM and FEM reproduce the broad high-efficiency angular plateau at 0° and 25° and the main spectral band near 490–498 nm. The reflective configuration exhibits a larger discrepancy than the transmissive case. Because the continuously rotating director is represented by piecewise-uniform longitudinal slices, this deviation is consistent with residual longitudinal-discretization error when the director rotates appreciably within an individual slice.

### *D. Gradient validation and topology optimization*

BCM uses only standard matrix operations, including matrix multiplications, linear solves, and inversions, and can therefore be differentiated directly. We test the resulting gradient near a resonance, where the transmission changes rapidly with geometry. The material system follows that of Section 3.A. Here, we additionally account for lithium-niobate dispersion [37]. The nanofin is replaced by the nanocylinder in Fig. 5(a), and the gradient of the transmittance with respect to the cylinder radius is evaluated over radii from 105 to 125 nm using 64-bit numerical precision. Central finite differences with a 0.005 nm step provide the reference, with EIG used to calculate the transmittances entering the finite-difference estimate and BCM used to obtain the automatic-differentiation gradient.

Fig. 5(b) shows the co-polarized transmission responses for the two orthogonal incident polarizations. The nanocylinder supports a resonance near a radius of 120 nm, providing a stringent gradient test because small geometrical perturbations produce rapid changes in the optical response. Despite this sharp variation, the automatic-differentiation gradients from BCM closely reproduce the finite-difference results, and the relative error remains below 1.1% throughout the scan, as shown in Figs. 5(c) and 5(d).

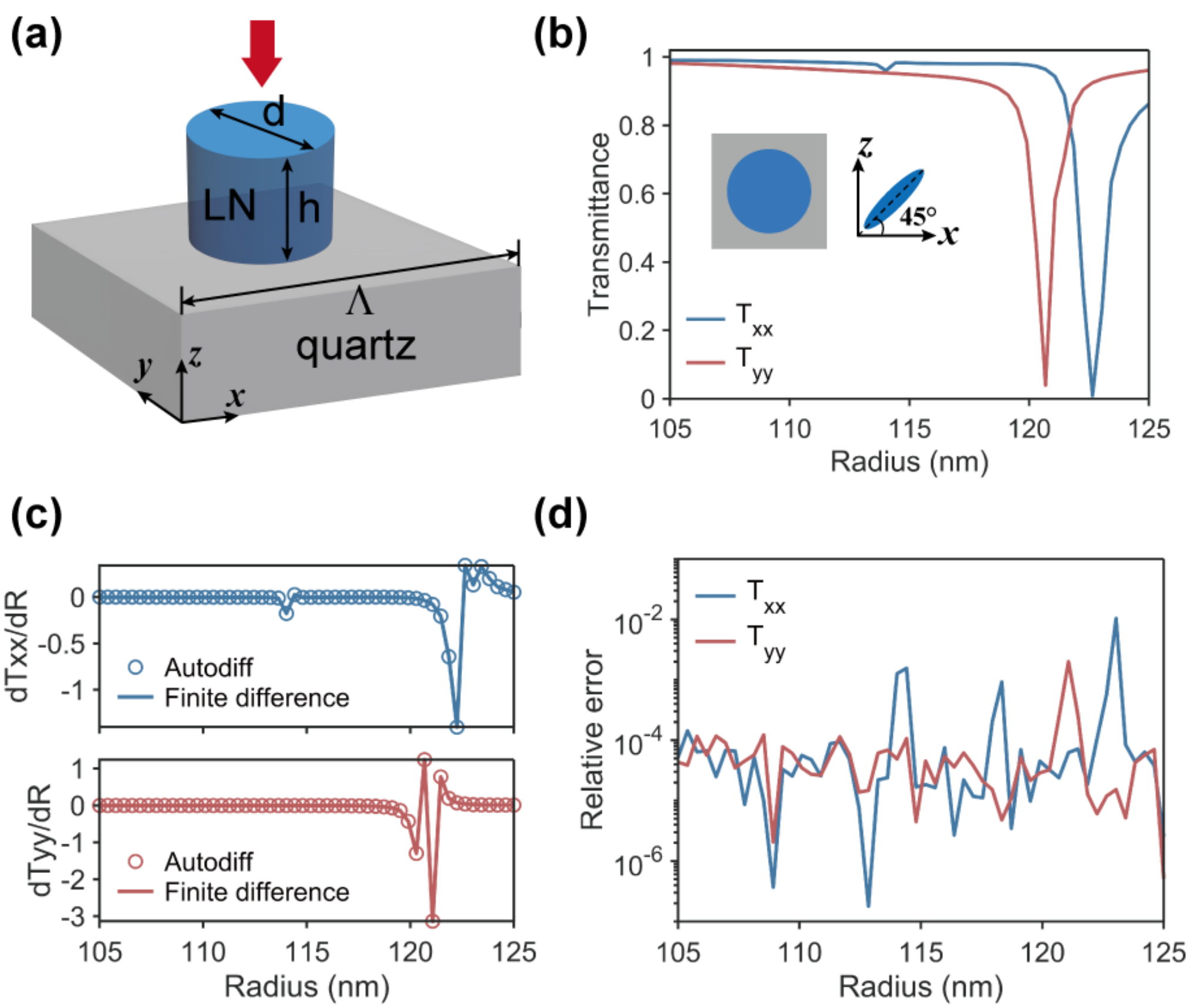

Fig. 5. Validation of automatic-differentiation gradients. (a) Lithium-niobate nanocylinder with a 340 nm period, 300 nm height, and fused-silica substrate. Normally incident light has a wavelength of 515 nm. (b) Co-polarized transmittance as the radius varies from 105 to 125 nm. (c) Radius derivatives from BCM automatic differentiation and central finite differences. (d) Relative gradient errors. A Fourier truncation of 361 harmonics is applied.

We next use the differentiable solver in a topology-optimization problem involving the intrinsic birefringence of lithium niobate. Previous work has shown that rotating the extraordinary axis of a lithium-niobate nanofin can couple two orthogonal resonant modes, producing elliptically polarized hybrid modes of opposite handedness and a strong circular-dichroism (CD) response [38]. CD is defined as $[\mathrm{T}_{RCP} - \mathrm{T}_{LCP}]/[\mathrm{T}_{RCP} + \mathrm{T}_{LCP}]$, where $\mathrm{T}_{LCP}$ and $\mathrm{T}_{RCP}$ are the co-polarized transmittances for left- and right-circularly polarized incidence, respectively. We first reproduce the previously reported co-polarized transmission spectra under normally incident x- and y-polarized light. As shown in Fig. 6(a), the calculated spectra recover the nearly degenerate resonance dips near 518 nm and their reported line shapes.

Our goal is to identify a geometrically $\mathrm{C}_{2\mathrm{v}}$-symmetric meta-atom with a strong CD response. Such a response requires controlled coupling between two orthogonal resonances and therefore depends jointly on the topology, optical-axis orientation, and operating wavelength. We optimize these variables simultaneously. The material system is the same as above, except that the lithium-niobate optical axis is rotated within the $xy$ plane. The permittivity distribution is

$$\boldsymbol{\varepsilon}(x, y; \zeta, \lambda) = \rho(x, y)\boldsymbol{\varepsilon}_{LN}(\zeta, \lambda) + [1 - \rho(x, y)]\boldsymbol{\varepsilon}_{air}, \tag{15}$$

where $\rho(x, y)$ is the lithium-niobate filling factor, $\zeta$ is the in-plane extraordinary-axis angle, and $\lambda$ is the operating wavelength. For topology-optimization, BCM evaluates the optical response and provides the corresponding gradients through automatic differentiation. The figure of merit (FoM) favors a CD approaching $-1$ while maintaining finite transmission and promoting nearly binary structures. The details of the optimization can be found in Supplementary Material D.

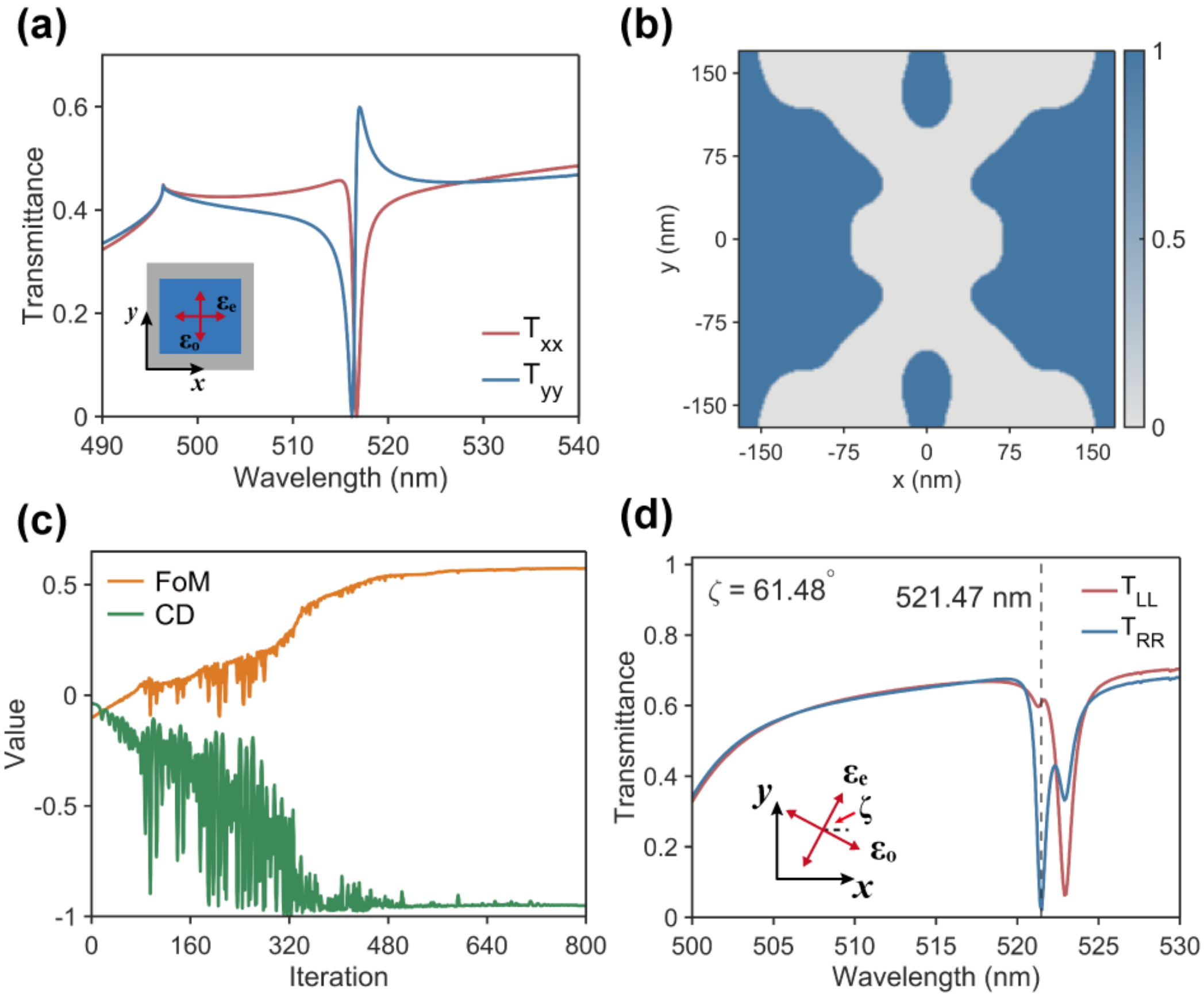

Fig. 6. Joint optimization of topology, wavelength, and optical-axis orientation. (a) Co-polarized transmission spectra of the uncoupled reference nanofin from Ref. [38]. (b) Optimized geometrically $C_{2v}$-symmetric meta-atom. Blue denotes lithium niobate, and gray denotes air. (c) Circular dichroism and objective value during 800 iterations. (d) Co-polarized transmission spectra for left- and right-circularly polarized incidence at the final axis orientation.

Starting from a random topology with $\lambda_0 = 522.0$ nm and $\zeta_0 = 24°$, BCM completes 800 optimization iterations in 550 s, achieving an approximately 16.4-fold speedup over the corresponding EIG-based optimization, which requires approximately 9000 s under the same optimization protocol and hardware configuration. The resulting geometrically $C_{2v}$-symmetric topology is shown in Fig. 6(b). A one-pixel-wide transition region is retained at the material boundary to approximate fabrication-induced edge smoothing. Fig. 6(c) illustrates that the CD fluctuates strongly during the 800 iterations, consistent with the sensitivity of coupled resonances to small structural changes. The optimization converges to $\lambda = 521.47$ nm and $\zeta = 61.48°$ with $\mathrm{CD} = -0.9503$, as shown in Fig. 6(d).

The optimization is sensitive to initialization: some random starting points remain near zero CD rather than converging to a strongly chiral structure. This behavior is consistent with a nonconvex, resonance-sensitive design landscape. The successful optimization demonstrates that gradients can be propagated through repeated full-tensor electromagnetic solves. Together with the gradient validation, this supports the use of BCM as a differentiable solver for inverse design of anisotropic periodic photonic structures.

## 4. DISCUSSION

The results show that the proposed formulation provides an eigendecomposition-free route to full-tensor RCWA while retaining the accuracy and numerical stability required for forward simulation and gradient-based inverse design. The mixed boundary representation contributes directly to this stability. The representation has an S-matrix-like structure, in which incoming and outgoing field components are distributed between the two ends of each interval instead of propagating the complete field state from one boundary to the other. Exponentially growing and decaying channels are therefore not explicitly accumulated as in a conventional transfer formulation, reducing the numerical imbalance that can arise in optically thick or strongly evanescent systems. Recursive interval doubling can therefore extend a locally accurate boundary relation to the full layer while preserving a stable interface-to-interface description.

The fourth-order Taylor expansion used here serves only as a local initializer and is not intrinsic to BCM. Higher-order polynomial or rational approximations, such as Padé-type constructions [31], could reduce the BCM order required for a prescribed accuracy. The accuracy of the overall RCWA solver is also distinct from the BCM propagation error. Fourier truncation introduces transverse-discretization error whose convergence depends strongly on the Fourier factorization, particularly at discontinuous material interfaces [15]. Li's factorization rules are therefore complementary to the present formulation: BCM governs longitudinal propagation after Fourier discretization, whereas Fourier factorization determines how material-field products are represented in the truncated Fourier basis. Increasing the BCM order cannot compensate for an insufficient Fourier basis or an inappropriate factorization rule.

The practical scope of BCM is best understood within the class of structures for which RCWA is naturally suited: media that are periodic in the transverse plane and piecewise layered, or can be accurately sliced, along the longitudinal direction. BCM is not uniformly advantageous across this class. At low Fourier truncation orders, its cascade overhead may exceed the cost of eigendecomposition, while eigendecomposition-based formulations remain preferable when propagation constants, eigenpolarizations, or internal modal compositions are themselves of primary interest. Its advantage becomes more pronounced when convergence requires a large Fourier basis, as can occur for TM-dominated responses or high-index-contrast structures [39]. In this regime, eigendecomposition becomes an increasingly significant computational bottleneck, making the boundary-cascade formulation comparatively more favorable. BCM is therefore particularly well suited to strongly coupled full-tensor anisotropic systems and to repeated-solve applications, including parameter sweeps, gradient-based inverse design, topology optimization, and data generation.

## 5. CONCLUSION

This work establishes a boundary-value cascade formulation for full-tensor anisotropic RCWA. The framework replaces the internal modal basis with interface-to-interface boundary field, retaining vectorial coupling while avoiding eigendecomposition and mode classification. It brings anisotropic electromagnetics into a form suited to automatic differentiation, GPU execution, and inverse design. Efficient access to full-wave electromagnetic responses and their gradients may further facilitate physics-informed surrogate modeling and large-scale data generation. With extensions to more general constitutive responses [40], BCM could provide a unified computational foundation for automated forward modeling and inverse design across a broader class of anisotropic photonic systems.

**Funding.** This work was supported by Scientific Research Innovation Capability Support Project for Young Faculty ZYGXQNJSKYCXNLZCXM-E8.

**Disclosures.** The authors declare no conflicts of interest.

**Data availability.** Data underlying the results presented in this paper are available from the authors upon reasonable request. The source code supporting the findings of this study will be made publicly available upon publication of the article.

**Supplemental document.** See Supplement for supporting content.